# Terahertz multispectral polarimetric imaging based on intensity measurement

Redwan Ahmad[1], Charles Simard[1], Rejeena R Sebastian[1], Jonathan Lafreniere-Greig[1], Xavier Ropagnol[1,2], and François Blanchard[1,*]

[1]Département de génie électrique, École de technologie supérieure (ÉTS), Montréal QC H3C 1K3, Canada
[2]Institut national de la recherche scientifique, EMT research center, Varennes QC J3X 1P7, Canada
*francois.blanchard@etsmtl.ca

**Abstract:** A compact terahertz (THz) polarimetric spectrometer and imaging system is demonstrated using polarization-sensitive frequency-selective surfaces (PS-FSS) and rapid, intensity-based detection. Real-time Stokes parameter extraction enables angle-of-linear-polarization (AoLP) measurements and quantitative retrieval of the birefringence of a 1-mm-thick *x*-cut quartz crystal, used here as a benchmark anisotropic sample. The extracted birefringence is in good agreement with THz time domain spectroscopy-based measurements, validating the accuracy of the proposed THz polarimetric approach. Signal-to-noise ratios (SNR) up to 87 dB across 0.23–0.41 THz ensure reliable discrimination of the measured spectral components above the noise floor. Extending this approach to raster-scanned THz imaging using distinct PS-FSS orientations (0°, 45°, 90°, and 135°) enables simultaneous mapping of frequency-dependent Stokes parameters. The resulting degree of linear polarization (DoLP) and AoLP maps exhibit well-resolved polarimetric contrast, consistent with numerical simulations and independent visible polarimetric measurements. Operating entirely without field-resolved detection or mechanical delay stages, the system provides a robust and compact platform for THz polarimetric spectroscopy and imaging of anisotropic materials.

## 1. Introduction

Polarimetric imaging has emerged as a powerful tool for material characterization, enabling the detection of anisotropy, stress-induced birefringence, and molecular orientation through polarization-resolved measurements [1]. In the visible and near-infrared (NIR) spectrum, compact polarimetric cameras are widely available, typically employing micro-polarizer arrays in which each super-pixel consists of four sub-pixels oriented at 0°, 45°, 90°, and 135° [2]. This configuration enables the simultaneous acquisition of polarization-resolved intensities, as shown in Fig. 1(a), allowing reconstruction of the Stokes parameters ($S_0$, $S_1$, $S_2$) and derived quantities such as the degree of linear polarization (DoLP) and angle of linear polarization (AoLP) [3-4]. Such systems have found extensive applications in biomedical imaging, surface inspection, and industrial quality control by revealing microstructural variations and surface textures that are not accessible with conventional intensity-only imaging [5-7]. Recent advances in visible- and infrared-wavelength polarimetric cameras, including metasurface-based Stokes imaging polarimeters, further highlight the progressive development of polarization-resolved imaging technologies, supported by earlier developments and subsequent extensions of metasurface concepts into the mid-infrared [8-10]. Despite their versatility, visible and infrared polarimetric systems are fundamentally limited by shallow penetration depths in dielectric and scattering materials. In contrast, terahertz (THz) radiation offers enhanced penetration into non-metallic materials such as plastics, wood, ceramics, and paper [11-12], positioning THz polarization-resolved characterization and imaging as a promising complementary modality for mapping polarization-dependent contrast in a wide range of industrial and scientific samples [13-19].

A wide range of THz polarimetric imaging techniques, predominantly based on THz time-domain spectroscopy (THz-TDS), have been developed to probe material anisotropy and enhance imaging contrast [20–27]. Early demonstrations employed ZnTe-based electro-optic polarization imaging to achieve full electric-field reconstruction and birefringence mapping [20], later extended through Jones-matrix analysis and balanced electro-optic detection to improve sensitivity and spatial resolution [21,22]. While these coherent THz-TDS approaches provide effective polarization retrieval, they inherently rely on mechanical delay lines, resulting in bulky and complex systems that are poorly suited for compact or real-time operation. To reduce system complexity, alternative strategies have been explored, including polarization-resolved single-pixel THz-TDS employing Hadamard coding [23], portable reflection-based scanners measuring orthogonal polarization states [24], and Monte Carlo-based approaches developed for biomedical applications [25,26]. However, these approaches typically require computationally intensive reconstruction, which limits real-time performance. Polarization-sensitive metasurface absorbers that encode THz polarization states into infrared thermal images have also been demonstrated [27], although this method is inherently limited by indirect readout mechanisms and slow response times. Collectively, these studies demonstrate the feasibility of THz polarimetric imaging for revealing polarization-dependent material contrast and anisotropy beyond conventional amplitude-based THz imaging [28], yet remain largely dependent on coherent detection or indirect reconstruction. More recently, intensity-based THz systems employing frequency-selective surfaces (FSS) have enabled real-time multispectral measurements without mechanical delay lines [29,30]; however, these implementations are limited to spectral intensity detection and lack polarization sensitivity. The integration of polarization-sensitive frequency-selective surfaces (PS-FSS) offers a promising route toward simultaneous frequency- and polarization-selective THz detection in the intensity domain [31], yet multispectral THz polarimetric imaging based solely on intensity-only detection has not been demonstrated to date.

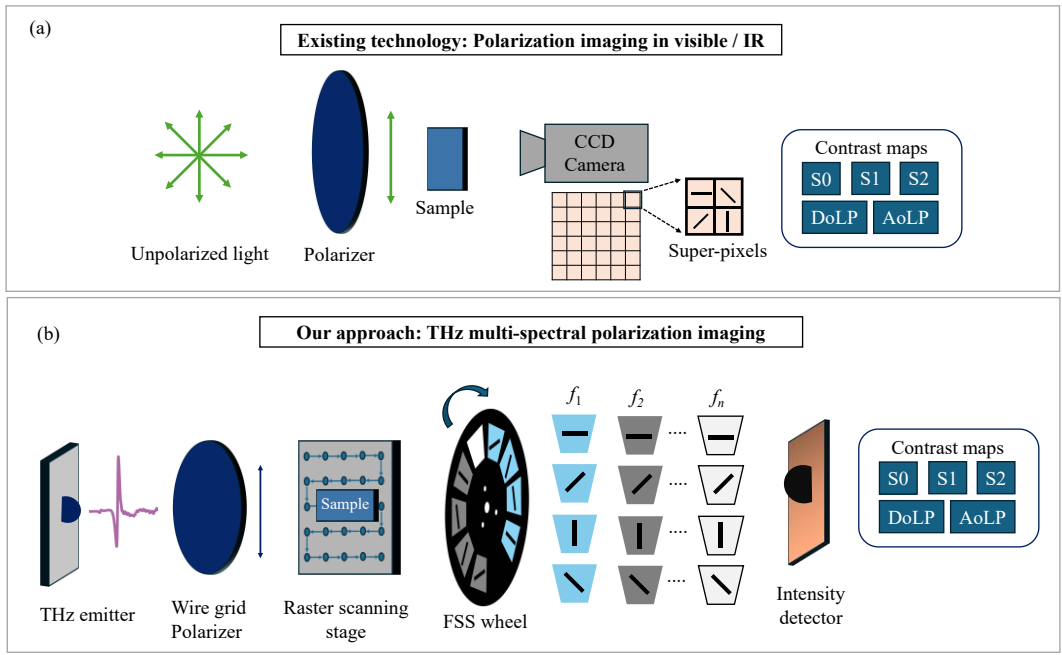

Fig. 1. Comparison of operating principles between (a) conventional polarimetric imaging at visible and near-infrared wavelengths, and (b) the proposed intensity-based multispectral THz polarimetric imaging approach.

In this work, we present a compact, intensity-only THz polarimetric spectrometer (TPS) that enables multispectral extraction of the Stokes parameters without coherent detection or mechanical delay stages. As illustrated in Fig. 1(b), the system integrates a set of frequency-selective polarizers into a metallic chopper wheel comprising sixteen sectors corresponding to four frequency bands and four analyzer orientations (0°, 45°, 90°, and 135°), allowing intensity-based retrieval of $S_0$, $S_1$, and $S_2$. The performance of the TPS is validated through polarization-resolved measurements of a 1 mm thick stacked quartz crystal, from which frequency-dependent Stokes parameters, DoLP, AoLP, and birefringence are quantitatively retrieved. We further demonstrate the extension of this approach to raster-scanned THz polarimetric imaging using double-layer stacked quartz crystals arranged in co-axial and cross-axial configurations, where reconstructed Stokes maps, DoLP, and AoLP images reveal distinct polarization contrasts associated with parallel and orthogonal optical-axis alignments. To provide independent validation, visible-wavelength Stokes imaging is performed on the same samples using a polarimetric CCD camera, and the agreement between the visible and THz results confirms the robustness of the proposed intensity-based multispectral polarimetric framework and illustrates the continuity of polarization-dependent material contrast from the visible to the THz regime. All experimental results are further supported by numerical simulations, showing consistent agreement with the measured polarization response across the investigated frequency range.

## 2. Polarization-sensitive FSS chopper wheel

To develop a multispectral polarimetric spectrometer, we integrate polarization discrimination and frequency selectivity within a single device. This is achieved using polarization-sensitive frequency-selective surface geometries, enabling selective transmission as a function of both polarization state and operating frequency. This behavior contrasts with that of conventional broadband polarizers and has been previously demonstrated in our earlier work [32].

Building on this established design, we implement PS-FSSs based on rectangular slot arrays, integrated into a rotating chopper wheel. A photograph of the fabricated device is shown in Fig. 2(a). The chopper wheel comprises sixteen PS-FSS sectors, formed by combining four distinct peak center frequencies with four polarization orientations (0°, 45°, 90°, and 135°). These orientations enable intensity-based retrieval of the Stokes parameters ($S_0$, $S_1$, and $S_2$) at each operating frequency. In addition, two sectors are reserved as open and blocked apertures to measure reference and noise signals, respectively. The PS-FSS chopper wheel was fabricated from 125 μm-thick stainless steel using laser micromachining (LPKF ProtoLaser U3). Microscopic images of the PS-FSS unit cells at polarization orientations of 0°, 45°, 90°, and 135° are shown in Fig. 2(b)(i–iv). The geometrical parameters of the rectangular slot array include the slot length $a$, slot width $b$, and the periodic spacings $c$ and $d$ along the horizontal and vertical directions, respectively.

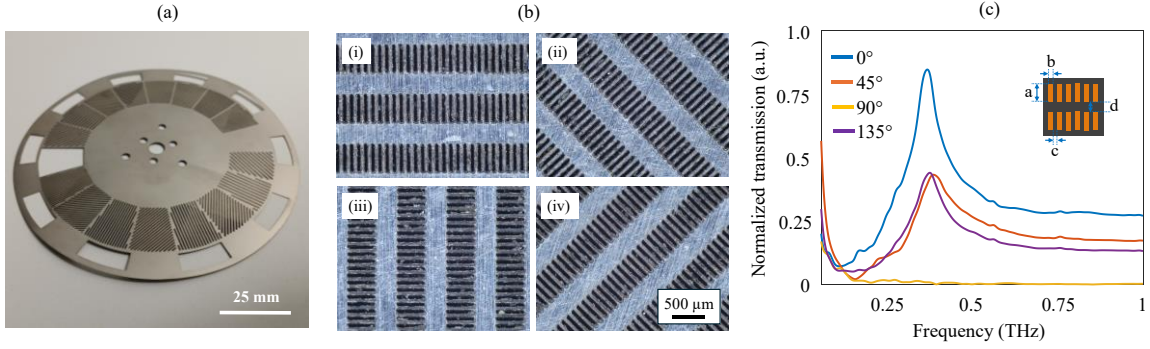

Fig. 2. (a) Photograph of fabricated PS-FSS chopper wheel; (b) microscopic images of PS-FSS of 0°, 45°, 90° and 135° orientation respectively; and (c) normalized transmission of PS-FSS ($a$=440 µm, $b$=$c$=55 µm, and $d$= 330 µm) for horizontal (0°), 45°, vertical (90°) and 135° polarization. Inset figure demonstrates the schematic of PS-FSS with representation of design parameters ($a$, $b$, $c$, and $d$).

We used TOPTICA TeraFlash Pro, a commercially available THz-TDS system, to experimentally characterize the PS-FSS in terms of peak center frequency and normalized transmission. Normalized transmission (NT) could be calculated as follows [33]:

$$NT = \frac{T_{sample}}{T_{ref}}, \qquad (1)$$

where $T_{sample}$ and $T_{ref}$ are the spectral amplitudes of the analyzed sample and without the sample (air) as a reference, respectively.

Figure 2(c) shows the normalized transmission, referenced to air, for polarization angles of 0°, 45°, 90°, and 135° for a representative PS-FSS with slot length $a$ = 440 µm, slot width $b$ = 55 µm, horizontal gap $c$ = 55 µm, and vertical gap $d$ = 330 µm. For 0° polarization, where the incident electric field is perpendicular to the long axis of the rectangular slots, the structure exhibits a band-pass response with a peak transmission at a center frequency $f_c$=0.37 THz. In addition to the dominant passband, a weak residual transmission is observed at higher frequencies. For vertical polarization (90°), the PS-FSS suppresses transmission to approximately 1% across a broad spectral range, demonstrating strong polarization discrimination. At intermediate polarization angles (45° and 135°), the transmission spectra are nearly identical, with peak transmission reduced to approximately 50% of that observed at 0°.

Notably, while the transmitted amplitude at the center frequency varies with polarization angle according to Malus' law, the peak center frequency remains invariant with respect to rotation. This behavior indicates that polarization rotation modulates the transmission amplitude without altering the resonant response of the PS-FSS, confirming its operation as a frequency-selective polarizer with high polarization selectivity. To enable multispectral operation, three additional PS-FSS designs were implemented by varying the slot length (700 µm, 500 µm, and 400 µm) and proportionally scaling the slot width and periodic gaps, yielding center frequencies of 0.23 THz, 0.33 THz, and 0.41 THz, respectively. Each PS-FSS was integrated into the rotatable chopper wheel at four polarization orientations (0°, 45°, 90°, and 135°).

## 3. Frequency-selective THz polarimetric system

Figure 3 provides an overview of the experimental setup, signal-processing workflow, and measurement configurations of the proposed frequency-selective THz polarimetric system. As shown in Fig. 3(a), THz radiation is generated using a photoconductive antenna (PCA) (THz-P-Tx: RTP220468B, TOPTICA Photonics) driven by a femtosecond fiber laser operating at 1560 nm wavelength, with an 80 fs pulse duration, 80 MHz repetition rate, and 30 mW average optical power. The emitted THz beam is collected and focused onto the sample using two off-axis parabolic mirrors ($M$1 and $M$2). The incident polarization is fixed to horizontal and purified using a wire-grid polarizer placed immediately before the sample. The transmitted signal is detected using a Schottky diode detector (3DL 12C LS2500 A2 from ACST), with a spherical silicon lens used to improve coupling efficiency. The combined PCA emission spectrum and detector responsivity fully cover the four operating frequency bands of the PS-FSSs (Appendix A, Fig. A1).

The detected THz signal is amplified using a Mini-Circuits ZFL-2000GH+ broadband amplifier and digitized with a 14-bit Acqiris SA240P high-speed acquisition card operating at 4 GS/s. This high sampling rate enables the extraction of the periodic response at the 80 MHz modulation frequency and its harmonics well above the low-frequency noise floor. As illustrated in Fig. 3(b), time-domain voltage traces are transformed into the frequency domain using a fast Fourier transform (FFT), yielding a comb of spectral peaks at the repetition frequency and its harmonics. Harmonic amplitudes exceeding a predefined noise threshold are retained and summed to enhance the signal-to-noise ratio (SNR), producing normalized intensity values for each PS-FSS slot and analyzer orientation. These intensity channels ($I_0$, $I_{45}$, $I_{90}$, and $I_{135}$) are used to compute frequency-dependent Stokes parameters ($S_0$, $S_1$, and $S_2$) and derived polarization metrics, including the degree and angle of linear polarization (DoLP and AoLP).

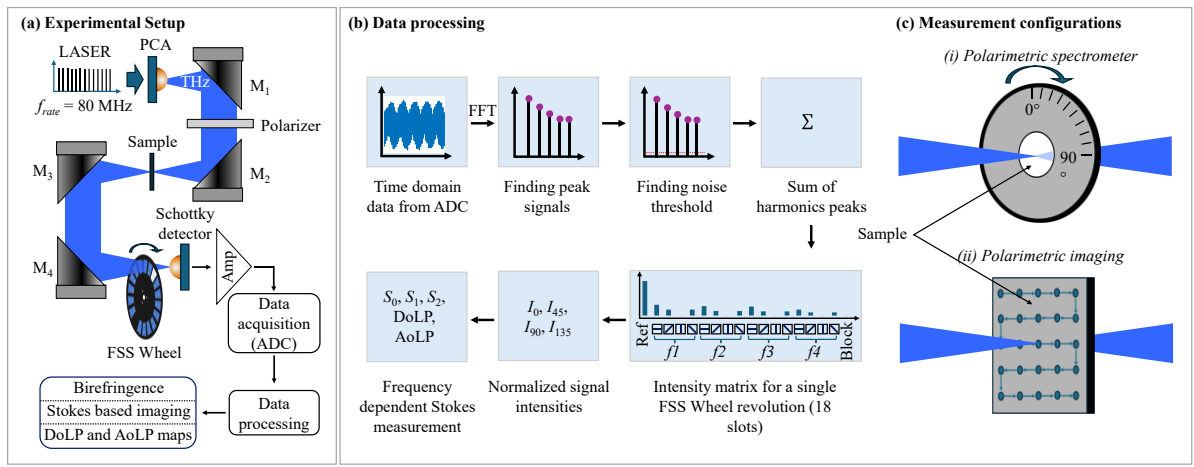

Fig. 3. (a) Schematic of the experimental setup; (b) data acquisition and processing module, illustrating real-time FFT analysis used to extract polarization-dependent spectral amplitudes; (c) system configurations: (i) polarimetric spectrometer and (ii) polarimetric imaging system, showing raster-scanning implementation for spatially resolved Stokes parameter retrieval and the corresponding degree and angle of linear polarization (DoLP and AoLP) mapping.

As shown in Fig. 3(c), the system operates in two measurement configurations. In the polarimetric spectrometer mode, the sample is mounted on a rotation stage at the focal plane to measure polarization-resolved transmission as a function of orientation. In the polarimetric imaging mode, the sample is raster-scanned using a motorized XY translation stage (Zaber X-MCB2), enabling spatially resolved acquisition of the Stokes parameters and corresponding DoLP and AoLP maps. In both configurations, a rotating PS-FSS chopper wheel positioned directly in front of the detector provides simultaneous multispectral and polarization-resolved modulation. The 4-inch diameter wheel, driven by a Thorlabs MC2000B controller, provides four analyzer orientations (0°, 45°, 90°, and 135°) at each frequency band, enabling real-time polarimetric readout within a single wheel rotation.

### 3.1 Spectral extraction and SNR analysis

Figure 4 illustrates the signal-processing performance and noise characteristics of the detection scheme. Figure 4(a) shows a representative time-domain voltage trace acquired over a single PS-FSS slot, along with a zoomed-in view highlighting the periodic modulation. The corresponding FFT magnitude spectrum (Fig. 4(b)) exhibits discrete harmonic peaks at integer multiples of the 80 MHz repetition frequency, extending up to 2 GHz. Noise performance was evaluated by varying the number of averaged acquisitions. As shown in Fig. 4(c), the mean noise level remains approximately constant at ~24 counts, while the standard deviation decreases from ~10 counts at $N=2$ to ~6 counts at $N=10$, indicating a stable baseline and consistent performance of the Schottky–amplifier–ADC detection chain. The resulting noise floor for ten averages is shown in Fig. 4(d).

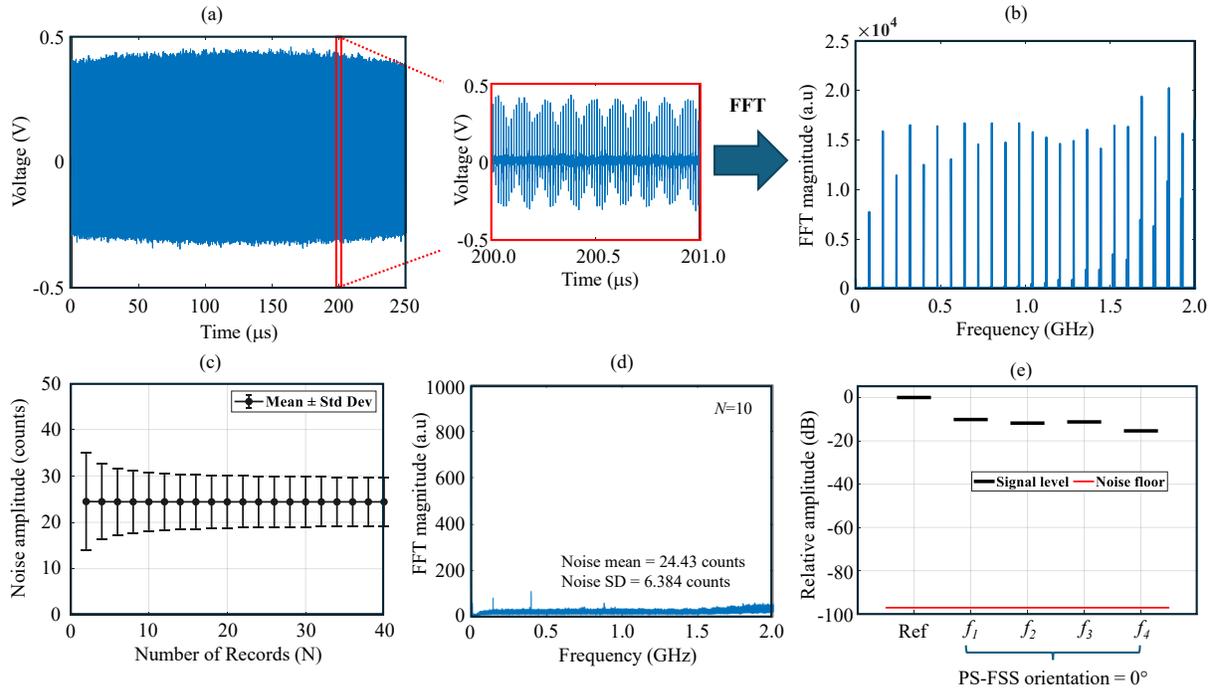

Fig. 4. (a) Representative time-domain voltage waveform and (b) corresponding FFT magnitude spectrum of the reference slot; (c) mean value and standard deviation of the noise level as a function of the number of averaged records, demonstrating the stability of the acquisition system; (d) FFT magnitude noise floor measured in the absence of a THz signal; (e) amplitude contrast for the reference slot and for the four operating center frequencies ($f_1$=0.23 THz, $f_2$=0.33 THz, $f_3$=0.37 THz, and $f_4$=0.41 THz).

To quantify signal visibility above the noise floor, the amplitude contrast (AC) is defined as

$$AC = 20\log_{10}\frac{\sum_{i=1}^{M}(A_i - \mu_{noise_i})}{\sigma_{noise}}, \qquad (2)$$

Where $A_i$ denotes the amplitude of the $i$-th harmonic component, $M$ is the number of harmonics considered, and $\mu_{noise}$ and $\sigma_{noise}$ represent the mean and standard deviation (SD) of the noise floor, respectively.

Figure 4(e) shows the amplitude contrast for the reference slot and the four PS-FSS frequency bands. The reference slot exhibits a maximum contrast of 97 dB, while PS-FSS slots at 0° orientation yield contrast values between 81 dB and 87 dB across the 0.23–0.41 THz range, demonstrating high sensitivity and excellent measurement repeatability.

Although the system does not directly measure the temporal electric-field waveform, the intensity-modulated response provided by the PS-FSS wheel enables robust reconstruction of polarization-dependent transmission as a function of frequency and orientation. This approach enables real-time multispectral THz polarimetric spectroscopy and imaging without the complexity of coherent THz time-domain systems.

## 4. Polarimetric spectrometer

To evaluate the functionality of the proposed polarimetric spectrometer, an $x$-cut quartz crystal was investigated through both simulation and experiment. Although the system employs a single fixed input polarization state, polarization analysis is achieved using four discrete analyzer orientations (0°, 45°, 90°, and 135°), while the detector records intensity only. Under these conditions, the Jones-matrix formalism provides a physically rigorous description of the phase retardation between the ordinary and extraordinary axes of the birefringent medium, governing the evolution of AoLP as the sample is rotated. This framework is therefore well-suited for predicting polarization-dependent observables from intensity-only, multispectral measurements obtained using the rotating PS-FSS wheel and Schottky diode detector. Details of the simulation model are provided in appendix B.

### 4.1 Stokes parameter reconstruction

The polarization state of the transmitted THz wave is described using the Stokes formalism. The Stokes parameters $S_0$, $S_1$, and $S_2$ were computed from the measured intensities at the four analyzer orientations according to standard definitions [4]:

$$S_0 = I_0 + I_{90} \qquad (3)$$

$$S_1 = I_0 - I_{90} \qquad (4)$$

$$S_2 = I_{45} - I_{135} \qquad (5)$$

where $I_0$, $I_{45}$, $I_{90}$ and $I_{135}$ denote the measured intensities at analyzer angles of 0°, 45°, 90°, and 135°, respectively. Since the system is sensitive only to linear polarization, the circular polarization component $S_3$ cannot be retrieved. From the Stokes parameters, the degree and angle of linear polarization are defined as [4]:

$$DoLP = \frac{\sqrt{S_1^2 + S_2^2}}{S_0} \tag{6}$$

$$AoLP = \frac{1}{2} \tan^{-1}\left(\frac{S_2}{S_1}\right) \tag{7}$$

**4.2 Simulation results**

In the simulations, the sample rotation angle $\theta_s$ was varied from 0° to 90°, while the birefringence $\Delta n$ was swept from 0.02 to 0.1 for a 1-mm-thick sample at a center frequency of 0.41 THz. As shown in Fig. 5(a), increasing birefringence leads to a systematic increase in the peak-to-peak AoLP variation ($\Delta_{AoLP}$), with two extrema appearing near 22.5° and 67.5°. The corresponding frequency-dependent behavior of $\Delta_{AoLP}$ over the 0.2–0.45 THz range is shown in Fig. 5(b), confirming that AoLP evolves jointly with frequency and birefringence due to frequency-dependent phase retardation. The combined dependence of AoLP on operating frequency and sample rotation angle for a 1-mm-thick quartz sample is summarized in Fig. 5(c), where higher frequencies yield stronger AoLP modulation.

To further assess thickness effects, a comprehensive Jones-matrix simulation was performed for a fixed birefringence of $\Delta n=0.05$, while sweeping the sample rotation angle (0–90°), thickness (0.1–4 mm), and frequency (0.10–0.40 THz). The results (Fig. A2(i–iv)) show that increasing thickness produces proportionally larger phase retardation and stronger AoLP modulation, while higher frequencies shift the AoLP response smoothly across the full angular range. These trends highlight the coupled influence of thickness and frequency on THz birefringence measurements.

**4.3 Experimental validation**

Experimentally, a 1-mm-thick stacked quartz sample (15 mm × 15 mm) was mounted on a precision rotation stage and measured in 10° increments (Fig. 3(c)(i)). For each orientation, intensity measurements were recorded at four center frequencies (0.23, 0.33, 0.37, and 0.41 THz) and four analyzer orientations. Representative normalized intensity traces $I_0$, $I_{45}$, $I_{90}$, and $I_{135}$ at $f_c=0.37$ THz are shown in Fig. 5(d). Each channel exhibits the expected modulation arising from the evolving projection of the rotated polarization state onto the analyzer axes. In particular, the $I_{45}$ and $I_{135}$ traces intersect near 50% normalized intensity at approximately 45° intervals, consistent with the symmetry of ±45° analyzer projections.

Notably, the Schottky detector incorporates a broadband log-spiral antenna that is intrinsically sensitive to both orthogonal linear polarization components. This dual-polarization sensitivity results in a measurable modulation of the $I_{90}$ channel as the quartz plate is rotated, reflecting the combined influence of the evolving polarization state and detector response.

Using the measured intensity values, the Stokes parameters were calculated using Eqs. (3)–(5), normalized by $S_0$, and the AoLP was subsequently extracted using Eq. (7). As shown in Fig. 5(e), rotation of the sample produces a periodic AoLP modulation with two extrema of opposite sign over a 90° rotation period. The peak-to-peak AoLP variation increases monotonically with frequency (Fig. 5(f), left axis), in agreement with the simulation results.

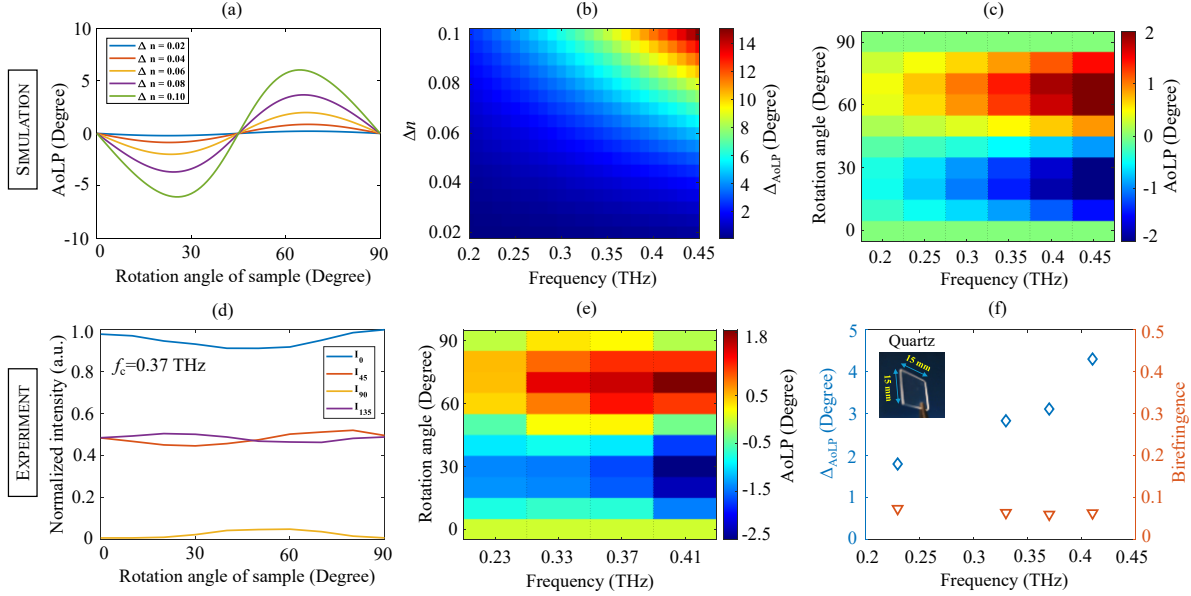

Fig. 5. Simulation results: (a) AoLP as a function of sample rotation angle for a 1-mm-thick sample with varying birefringence at 0.41 THz; (b) simulated AoLP variation as a function of frequency and birefringence for a 1-mm-thick sample; (c) simulated AoLP map as a function of operating frequency and sample rotation angle. Experimental results: (d) normalized intensities measured at analyzer orientations of 0°, 45°, 90°, and 135° at $f_c$=0.37 THz for 1-mm-thick stacked quartz sample; (e) measured AoLP at four center frequencies (0.23, 0.33, 0.37, and 0.41 THz) as a function of quartz rotation angle from 0° to 90°; (f) peak-to-peak AoLP variation at the four center frequencies (left axis) and the corresponding frequency-dependent birefringence of a 1-mm-thick stacked quartz sample (right axis).

### 4.4 Birefringence retrieval

To quantitatively extract the birefringence of the quartz sample, an optimization-based fitting procedure was implemented using the measured $\Delta_{AoLP}$ values at the four operating frequencies and the known sample thickness. A Jones-matrix forward model was used to simulate the AoLP response as a function of rotation angle, and a direct-search algorithm minimized the discrepancy between simulated and experimental $\Delta_{AoLP}$. The retrieved frequency-dependent birefringence values are shown in Fig. 5(f) (right axis), yielding an average birefringence of approximately $\Delta n \approx 0.06$ over the 0.23–0.41 THz range. These values are consistent with independent THz-TDS measurements and reported literature values [34]. A small residual discrepancy on the order of $\sim 5\times10^{-3}$ is attributed to alignment tolerances, finite polarization purity, and interface-induced depolarization effects.

Potential strategies to minimize or compensate for these offsets, including improved alignment procedures, the use of narrow band-pass filters, and depolarization correction, are briefly discussed in section 6 (Future Perspectives).

## 5. Polarimetric imaging

Subsequently, we performed a comparative investigation of Stokes-based imaging in both visible and THz ranges using intensity-only detection. In both cases, stacked quartz samples consisting of two 0.5 mm-thick plates were investigated. Two sample configurations were investigated based on the relative orientation of the birefringent crystal fast axes with respect to their slow axes: a co-axial configuration with aligned fast axes and a cross-axial configuration with orthogonal fast-axis orientations. Sample rotation was used to probe the resulting polarization-dependent behavior in both the visible and terahertz spectral domains.

### 5.1 Visible polarimetric camera

The experimental configuration of the visible polarimetric imaging system is shown in Fig. 6(a). A commercially available polarimetric CCD camera (Alkeria CELERA P series) was used to acquire Stokes-resolved images of the stacked quartz samples. Each super-pixel consists of four sub-pixels with integrated linear polarizers oriented at 0°, 45°, 90°, and 135°, enabling simultaneous acquisition of polarization-resolved intensities and reconstruction of the

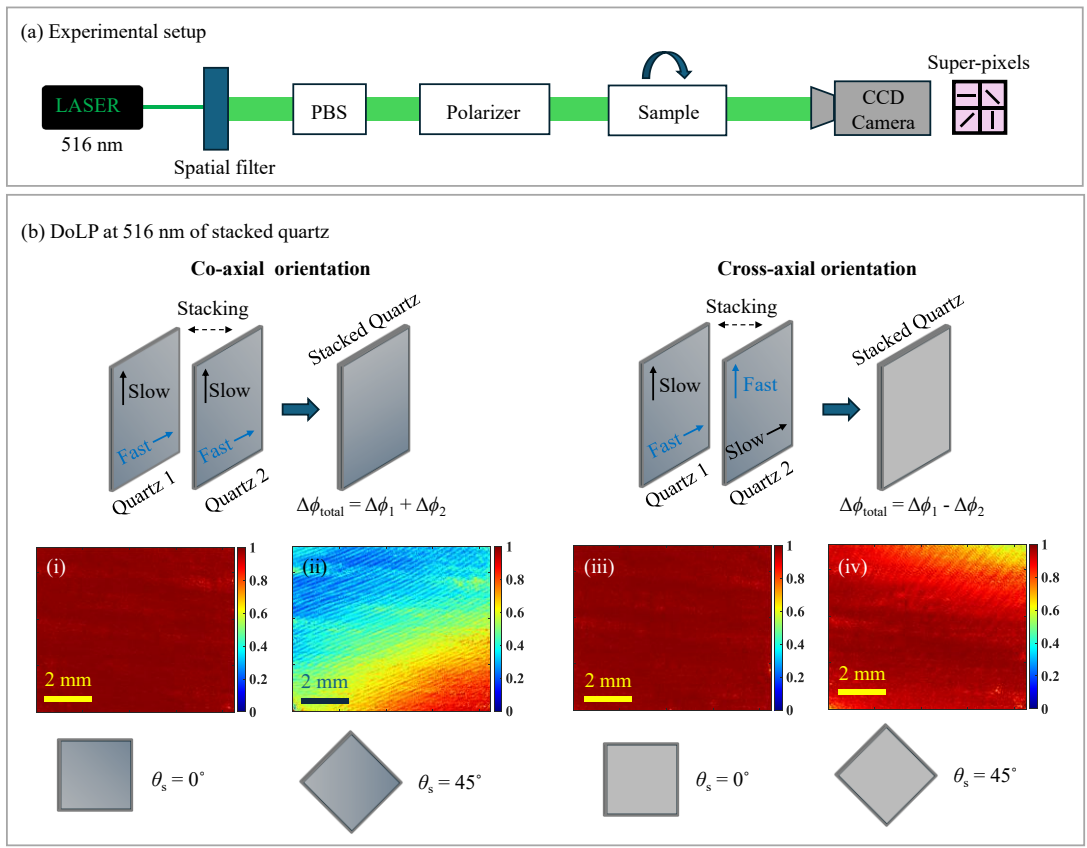

Fig. 6. (a) Experimental setup for visible-wavelength Stokes polarimetric imaging at 516 nm using a micro-polarizer CCD camera. PBS denotes a polarizing beam splitter. (b) Measured DoLP maps of stacked quartz for co-axial fast-axis orientation at analyzer angles of (i) 0° and (ii) 45°, and for cross-axial fast-axis orientation at (iii) 0° and (iv) 45°.

Stokes parameters, as well as derived quantities such as DoLP and AoLP [35]. The effective output resolution after super-pixel processing is 1232 × 1028 pixels.

The samples were illuminated using a continuous-wave 516 nm laser. A spatial filter was placed at the Fourier plane of the beam to ensure a uniform illumination profile, and a linear polarizer was positioned before the sample to define the incident polarization state. The sample was mounted on a rotation stage, allowing measurements at different azimuthal orientations $\theta_s$ to evaluate the angular dependence of the Stokes parameters. Due to the sample dimensions exceeding the active sensor area, the recorded images correspond to a portion of the sample.

Figure 6(b) presents the measured DoLP maps for co-axial and cross-axial stacking configurations at sample orientations of $\theta_s = 0°$ and 45°. For the co-axial configuration, high DoLP values approaching unity are observed at $\theta_s = 0°$, while a pronounced reduction in DoLP occurs at $\theta_s = 45°$, consistent with polarization transformation induced by accumulated birefringent phase retardation (Fig. 6(b)(i–ii)). In contrast, the cross-axial configuration exhibits consistently high DoLP across both orientations, reflecting partial cancellation of the birefringent phase retardation and confirming theoretical expectations (Fig. 6(b)(iii–iv)). These results demonstrate the sensitivity of Stokes-based imaging to the relative orientation of anisotropic layers.

The corresponding AoLP maps at 516 nm for the co-axial configuration at $\theta_s = 0°$ and 45° are shown in Fig. A3(a)(i–ii), while those for the cross-axial configuration are presented in Fig. A3(a)(iii–iv). The largest AoLP variations are observed for $\theta_s = 45°$, indicating enhanced interaction between the incident polarization and the birefringent axes of the stacked quartz. These experimental observations are further corroborated by numerical simulations shown in Fig. A3(b).

## 5.2 Multispectral polarimetric THz imaging

THz polarization-resolved imaging was subsequently performed at the selected center frequencies using the PS-FSS analyzer. The same stacked quartz sample was investigated, and raster scanning was carried out with a step size of 1 mm. Measurements were conducted for both co-axial and cross-axial stacking configurations. Representative Stokes $S_0$ maps acquired at 0.23 THz for sample orientations of $\theta_s = 0°$ and 45° under co-axial alignment are shown in Fig. 7(a)(i–ii), with the corresponding DoLP distributions presented in Fig. 7(a)(iv–v). The $S_0$ images primarily reflect the sample geometry and orientation, while the DoLP maps reveal pronounced polarization-dependent contrast. In particular, a reduction in DoLP at $\theta_s = 45°$ is observed for the co-axial configuration, consistent with birefringent phase retardation and in good agreement with numerical simulations (Fig. A4(a–b)).

A direct comparison between co-axial and cross-axial configurations, shown in Fig. 7(a)(ii–iii), further highlights this distinction. While the corresponding $S_0$ images appear similar, their DoLP maps (Fig. 7(a)(v–vi)) exhibit clear differences, demonstrating the ability of the proposed system to discriminate optical-axis alignment through polarization-sensitive contrast. The extracted AoLP maps for three representative cases, namely co-axial stacking at $\theta_s = 0°$ and 45°, and cross-axial stacking at $\theta_s = 45°$, are shown in Fig. 7(a)(vii–ix). Across the interior of the quartz crystal, the AoLP remains close to 0°, consistent with theoretical expectations and simulations (Fig. A4(c–d)), while localized deviations are observed near the sample edges due to depolarization and finite spatial resolution.

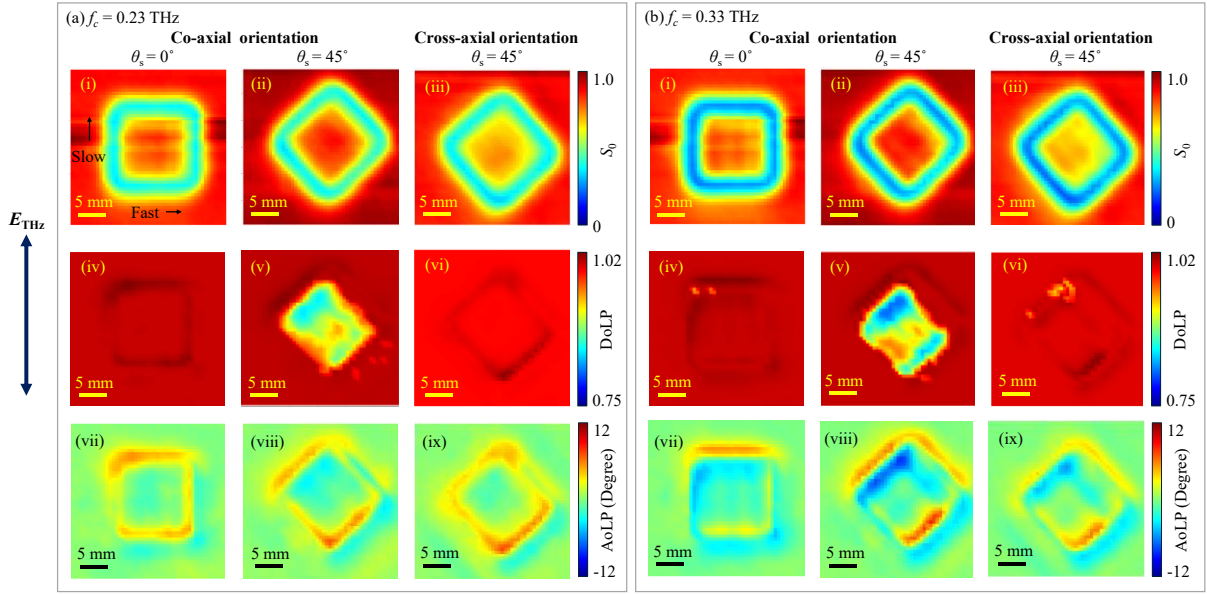

Fig. 7. Polarization-resolved THz images of the stacked quartz sample at (a) 0.23 THz and (b) 0.33 THz. Panels (i–iii) show the reconstructed Stokes $S_0$ for (i) 0° co-axial, (ii) 45° co-axial, and (iii) 45° cross-axial sample orientations. Panels (iv–vi) present the corresponding DoLP maps, while panels (vii–ix) display the extracted AoLP maps. The images reveal distinct polarization-dependent contrast between co-axial and cross-axial configurations, with localized reductions in DoLP observed near the sample edges.

Figure 7(b) presents corresponding polarization-resolved images acquired at 0.33 THz. Similar trends are observed at this frequency, with $S_0$ maps remaining largely invariant across configurations, while DoLP and AoLP maps continue to exhibit polarization-dependent contrast associated with birefringence and sample orientation. Comparable Stokes, DoLP, and AoLP images were also obtained at center frequencies of 0.37 THz and 0.41 THz (not shown), demonstrating the multispectral capability of the proposed THz polarimetric imaging system.

Although stacked quartz is employed here as a representative birefringent sample for proof-of-concept validation, the demonstrated methodology is broadly applicable to polarization-resolved THz imaging of anisotropic materials. By eliminating the need for coherent detection and mechanical delay stages, the proposed compact, intensity-based framework provides a practical route toward multispectral THz polarimetric imaging for material characterization and nondestructive inspection.

## 6. Future perspectives

Due to the wideband nature of the band-pass filters employed in the present system, the extracted birefringence represents a weighted average over the filter bandwidth rather than a strictly monochromatic value at the center frequency. Future implementations could improve frequency specificity by employing narrower band-pass filters with sharper spectral roll-off and by increasing the number of operating frequency channels, thereby reducing dispersion-induced averaging effects and enhancing measurement accuracy.

In intensity-based, frequency-selective THz systems using band-pass filters, Fabry–Pérot reflections are inherently suppressed for thin samples such as the 0.5–1 mm quartz crystals investigated here, owing to the weak amplitude of internal echoes. For thicker samples, however, multiple reflections may introduce measurable distortions. Incorporating Fabry–Pérot effects into the numerical modeling and signal processing framework represents a promising route to compensate for such artifacts and improve quantitative accuracy in future studies.

While the current system operates up to 0.41 THz, further optimization of the polarization-sensitive frequency-selective surface design could extend the accessible frequency range. The Schottky detector employed in this work remains operational up to approximately 1.0 THz, supporting the feasibility of broader spectral coverage. In addition, emerging CMOS-based THz receiver technologies offer a promising alternative for broadband applications [36], particularly where large-area integration and system scalability are desired.

Finally, extending the simulation framework to explicitly account for depolarization effects arising from complex field interactions, particularly at lower frequencies, could further improve birefringence estimation accuracy and interpretation of polarization-resolved THz images.

**7. Conclusion**

We have developed a compact, multispectral THz polarimetric spectrometer and imaging system based on frequency-selective polarizers integrated into a rotating chopper wheel, enabling fast, intensity-only polarization detection without coherent probe beams or mechanical delay lines. The proposed system enables retrieval of the birefringence of quartz in agreement with established literature values and demonstrates raster-scanned THz polarimetric imaging with distinct DoLP and AoLP contrast for co-axial and cross-axial crystal orientations. By reconstructing Stokes parameters directly from intensity measurements, this approach avoids phase-resolved detection while improving system compactness, robustness, and experimental simplicity. Overall, the demonstrated platform provides a promising pathway toward practical intensity-based THz polarimetric spectroscopy and imaging for material characterization and non-destructive evaluation.

**Appendix-A: THz emitter characterization and reference birefringence measurement**

The THz emitter based on a PCA was characterized using a commercial THz-TDS system (TOPTICA TeraFlash Pro), with both emission and detection performed using PCA modules. As shown in Fig. A1(a) (left axis), the emitter exhibits a broadband spectral response with a peak centered near 0.35 THz. The normalized responsivity of the Schottky diode detector, shown on the right axis of Fig. A1(a), reaches its maximum around 0.1 THz and gradually decreases toward higher frequencies.

In addition, the birefringence of the stacked quartz sample (total thickness of 1 mm) was independently determined using the same THz-TDS system as a reference measurement. From the time-domain data, the refractive index $n$ and birefringence $\Delta n$ were extracted according to Eqs. (8) and (9) [37]:

$$n(f) = 1 + \frac{c\Delta\phi}{2\pi f d} \tag{8}$$

$$\Delta n = n_e - n_o \tag{9}$$

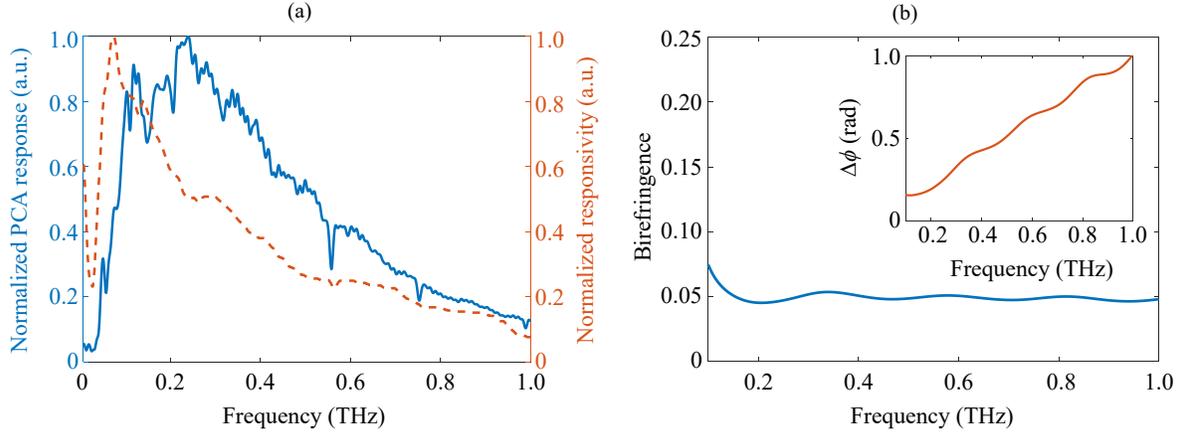

Fig. A1. (a) Normalized emission spectrum of the THz photoconductive antenna (left axis) and normalized responsivity of the Schottky diode detector (right axis), showing the broadband spectral overlap between the emitter output and detector sensitivity. (b) frequency-dependent birefringence of a 1 mm thick stacked quartz crystal obtained from THz-TDS. The inset shows the corresponding phase retardation as a function of frequency for the same sample.

where $c$ is the speed of light, $f$ is the frequency, and $d$ denotes the thickness of the birefringent sample. The terms $n_e$ and $n_o$ correspond to the extraordinary and ordinary refractive indices, respectively, and $\Delta\phi$ represents the phase retardation between orthogonal polarization components, defined as $\Delta\phi = \phi_x - \phi_y$.

Figure A1(b) shows the frequency-dependent birefringence of the $x$-cut stacked quartz crystal over the 0.1–1 THz range. The inset of Fig. A1(b) displays the corresponding phase retardation as a function of frequency for the same sample.

**Appendix-B: Jones-matrix model for polarization evolution in a birefringent sample**

The polarization evolution of a linearly polarized THz wave propagating through a birefringent material can be described using the Jones-matrix formalism. When the birefringent sample is rotated by an angle $\theta$ with respect to the laboratory reference frame, the effective Jones matrix of the system is given by [38,39]:

$$J_{rotated}(\theta) = R(\theta)\, J_{sample}\, R(-\theta) \qquad (10)$$

where $R(\theta)$ and $R(-\theta)$ are the rotation and inverse rotation matrices, respectively. These matrices are expressed as

$$R(\theta) = \begin{bmatrix} \cos\theta & -\sin\theta \\ \sin\theta & \cos\theta \end{bmatrix} \qquad (11)$$

$$R(-\theta) = \begin{bmatrix} \cos\theta & \sin\theta \\ -\sin\theta & \cos\theta \end{bmatrix} \qquad (12)$$

The Jones matrix of the birefringent sample itself is written as:

$$J_{Sample} = \begin{bmatrix} e^{i\frac{\Delta\phi}{2}} & 0 \\ 0 & e^{-i\frac{\Delta\phi}{2}} \end{bmatrix} \quad (13)$$

where $\Delta\phi$ is the birefringence-induced phase retardation between the ordinary and extraordinary polarization components. This phase retardation is given by:

$$\Delta\phi = \frac{2\pi f d \Delta n}{c} \quad (14)$$

where $f$ represents the frequency, $d$ is the sample thickness, $\Delta n$ denotes the birefringence, and $c$ corresponds to the speed of light.

The transmitted electric field after propagation through the rotated birefringent sample is obtained as [40,41]:

$$E_{out} = J_{rotated} E_{in} \quad (15)$$

where $E_{in}$ represents the incident electric field vector.

Polarization analysis is performed using a linear analyzer oriented at an angle $\alpha$. The corresponding Jones matrix is given by:

$$J_{analyzer}(\alpha) = \begin{bmatrix} \cos^2(\alpha) & \cos(\alpha)\sin(\alpha) \\ \cos(\alpha)\sin(\alpha) & \sin^2(\alpha) \end{bmatrix} \quad (16)$$

The electric field transmitted through the analyzer is therefore

$$E(\alpha) = J_{analyzer}(\alpha) E_{out} \quad (17)$$

and the detected intensity is calculated as

$$I(\alpha) = |E(\alpha)|^2 \quad (18)$$

**Appendix-C: Simulated AoLP response as a function of thickness and frequency**

To gain further insight into the evolution of birefringence-induced phase retardation, numerical simulations were performed to evaluate the AoLP response of a birefringent sample with a fixed birefringence of $\Delta n = 0.05$ over a broad range of sample thicknesses and operating frequencies. The resulting AoLP distributions, shown in Fig. A2(i–iv), illustrate how the AoLP depends on the accumulated phase difference between the ordinary and extraordinary field components. As both frequency and thickness increase, the phase retardation becomes more pronounced, leading to stronger AoLP variations as a function of sample rotation angle. These simulations demonstrate that even modest changes in birefringence or propagation length can produce measurable AoLP variations, highlighting the sensitivity of AoLP-based measurements to birefringent phase accumulation and supporting its use for qualitative assessment of polarization effects in anisotropic materials.

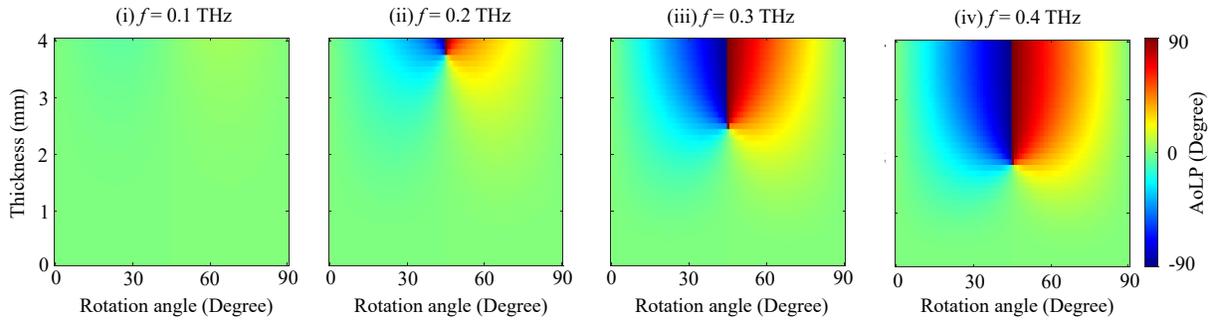

Fig. A2. Simulated AoLP as a function of sample thickness and rotation angle for distinct operating frequencies: (i) 0.1 THz, (ii) 0.2 THz, (iii) 0.3 THz, and (iv) 0.4 THz, illustrating the evolution of birefringence-induced phase retardation with both thickness and frequency.

**Appendix-D: Visible-wavelength AoLP imaging and simulation comparison**

AoLP images of a 1 mm thick stacked quartz sample were acquired at 516 nm using a polarimetric CCD camera, as shown in Fig. A3(a). As a reference, an initial AoLP image was recorded without the sample, yielding an AoLP value close to 0°. Measurements were then performed for both co-axial and cross-axial stacking configurations, with the sample oriented at $\theta_s = 0°$ and 45°.

For the co-axial configuration, a clear angular dependence of the AoLP is observed. At $\theta_s = 0°$, the measured AoLP remains close to 0°, whereas at $\theta_s = 45°$, the AoLP shifts toward higher values approaching 90°, reflecting the accumulation of birefringence-induced phase retardation in the aligned quartz plates. This behavior is characteristic of anisotropic birefringent media and indicates a strong polarization-modulating response.

In contrast, for the cross-axial configuration, the AoLP remains approximately constant across both rotation angles, indicating substantial suppression of the net birefringent effect due to the orthogonal alignment of the optical axes. This response is consistent with the expected cancellation of phase retardation in the cross-oriented stacking geometry.

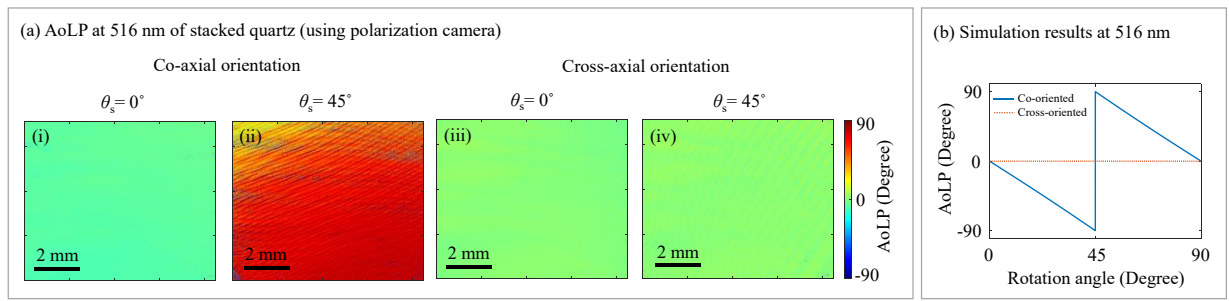

Fig. A3. (a) Experimentally obtained AoLP maps at 516 nm for the stacked quartz sample: (i–ii) co-axial orientation at 0°, and 45°, and (iii–iv) cross-axial orientation at 0°, and 45°. (b) Corresponding simulation results showing AoLP as a function of rotation angle for a 1 mm stacked quartz sample at 516 nm for both co-axial and cross-axial orientation.

Numerical simulations performed using a birefringence value of $\Delta n = 0.009$ [42] reproduce the main trends observed experimentally. The corresponding simulated AoLP response at 516 nm is shown in Fig. A3(b), demonstrating qualitative agreement with the experimental measurements and supporting the interpretation of the visible wavelength polarimetric results.

**Appendix-E: Simulated DoLP and AoLP dependence on birefringence and sample rotation**

Numerical simulations were performed to investigate the dependence of the DoLP and the AoLP on the sample rotation angle and birefringence at a center frequency of 0.41 THz. The simulations considered two stacked birefringent plates, each with a thickness of 0.5 mm, arranged in both co-axial and cross-axial configurations. The birefringence $\Delta n$ was varied from 0.02 to 0.1.

Figure A4(a) shows the simulated DoLP response for the co-axial configuration. A pronounced modulation of DoLP is observed as a function of rotation angle, with a minimum occurring near 45°. The depth of this modulation increases with increasing birefringence, indicating enhanced polarization sensitivity for stronger birefringent contrast. In contrast, for the cross-axial configuration shown in Fig. A4(b), the DoLP remains close to unity across the full range of rotation angles, reflecting effective compensation of birefringent phase retardation due to the orthogonal alignment of the optical axes.

The corresponding AoLP simulations are presented in Fig. A4(c–d). For the co-axial configuration (Fig. A4(c)), the AoLP exhibits a characteristic angular dependence, remaining close to 0° at rotation angles of 0°, 45°, and 90°, while showing pronounced extrema near 22.5° and 67.5°. The magnitude of these AoLP variations increases as the birefringence increases, consistent with stronger phase retardation. In contrast, for the cross-axial configuration (Fig. A4(d)), the AoLP remains approximately constant at 0° over all rotation angles and birefringence values, indicating suppression of net polarization rotation.

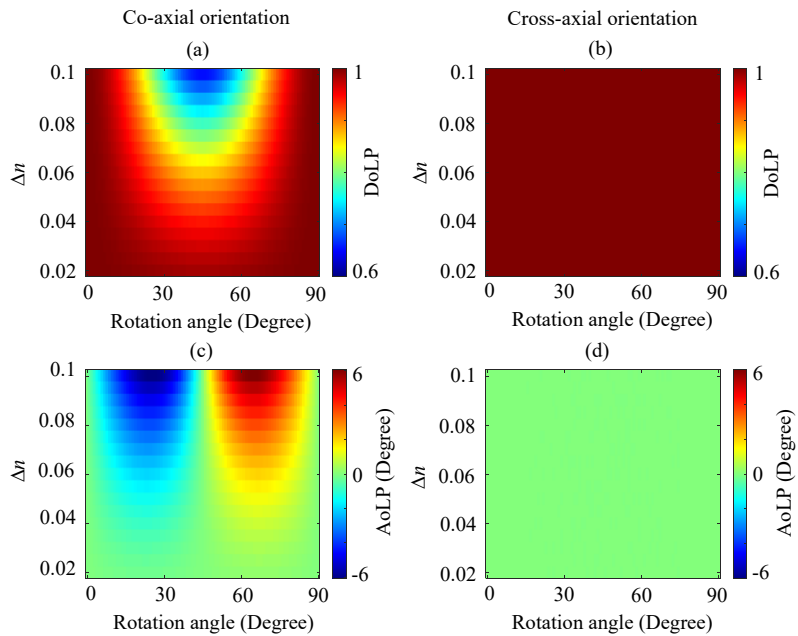

Fig. A4. Simulated DoLP as a function of sample rotation angle by varying birefringence in (a) co-axial and (b) cross-axial configurations of two 0.5-mm-thick birefringent sample. The corresponding AoLP responses as a function of rotation angle for different birefringence values are shown in (c) co-axial and (d) cross-axial configurations.

These simulations further support the interpretation of the experimental THz polarimetric imaging results by illustrating the distinct DoLP and AoLP signatures associated with co-axial and cross-axial birefringent stacking geometries.

**Funding.** This work was financially supported by the Natural Sciences and Engineering Research Council of Canada (NSERC, Grant No. 2023-03322) and the Canada Research Chairs Program (CRC-2024-00354). Redwan Ahmad acknowledges support from the PBEEE doctoral research fellowship (No. 319432) awarded by the Fonds de recherche du Québec – Nature et technologies (FRQNT).

**Acknowledgment.** The authors also extend their thanks to all members of the TeraÉTS Laboratory, ÉTS, for their insightful assistance and discussion.

**Disclosures.** The authors declare no conflicts of interest.

**Data availability**. Data underlying the results presented in this paper are not publicly available at this time but may be obtained from the authors upon reasonable request.

**References**

1. N. Baek, Y. Lee, T. Kim, *et al.*, "Lensless polarization camera for single-shot full-stokes imaging," APL Photonics **7**, 116107 (2022).
2. J. Liu, Z. Zhu, W. Li, *et al.*, "Polarized image fusion strategy based on multi-scale feature fusion," Opt. Express **33**(13), 27294–27305 (2025).
3. P.-J. Lapray, L. Gendre, A. Foulonneau, and L. Bigué, "An FPGA-based pipeline for micropolarizer array imaging," Int. J. Circuit Theory Appl. **46**(9), 1675–1689 (2018).
4. J. Raffoul, D. LeMaster, and K. Hirakawa, "Framework for improving DOLP and AOLP reconstruction quality in microgrid polarimeters," Opt. Express **30**(26), 48004–48020 (2022).
5. J. Zuo, J. Bai, S. Choi, A. Basiri, X. Chen, C. Wang, and Y. Yao, "Chip-integrated metasurface full-Stokes polarimetric imaging sensor," Light: Science & Applications **12**, 218 (2023).
6. L. Zhang, C. Zhou, B. Liu, Y. Ding, H.-J. Ahn, S. Chang, Y. Duan, M. T. Rahman, T. Xia, X. Chen, Z. Liu, and X. Ni, "Real-time machine learning–enhanced hyperspectro-polarimetric imaging via an encoding metasurface," Science Advances **10**, 36 (2024)
7. C. He, H. He, J. Chang, B. Chen, H. Ma, and M. J. Booth, "Polarisation optics for biomedical and clinical applications: a review," Light: Science & Applications **10**, 194 (2021)
8. Lisa W. Li, Jaewon Oh, Harris Miller, Federico Capasso, and Noah A. Rubin, "Flat, wide field-of-view imaging polarimeter," Optica **12**(6), 799-811 (2025).

9. H. Tan, J. Meng, and K. B. Crozier, "Multianalyte detection with metasurface-based midinfrared microspectrometer," ACS Sensors **9**(11), 5839–5847 (2024).
10. H. Zhao, Y. Li, G. Jia, *et al*., "Comparing analysis of multispectral and polarimetric imaging for mid-infrared detection blindness condition," Appl. Opt. **57**(24), 6840–6850 (2018).
11. A. Nakanishi and H. Satozono, "Terahertz optical properties of wood–plastic composites," Appl. Opt. **59**(4), 904–909 (2020).
12. S. Wietzke, C. Jansen, T. Jung, *et al*., "Terahertz time-domain spectroscopy as a tool to monitor the glass transition in polymers," Opt. Express **17**(21), 19006–19014 (2009).
13. K. Xu, E. Bayati, K. Oguchi *et al*., "Terahertz time-domain polarimetry (THz-TDP) based on the spinning E-O sampling technique: determination of precision and calibration," Opt. Express **28**(9), 13482–13496 (2020).
14. C. M. Morris, R. V. Aguilar, A. V. Stier, and N. P. Armitage, "Polarization modulation time-domain terahertz polarimetry," Opt. Express **20**(11), 12303–12317 (2012).
15. X. Chai, X. Ropagnol, L. S. Mora, S. M. Raeiszadeh, S. Safavi-Naeini, F. Blanchard, and T. Ozaki, "Stokes–Mueller method for comprehensive characterization of coherent terahertz waves," Scientific Reports **10**, 15426 (2020).
16. S.-X. Huang, G.-B. Wu, B.-J. Chen, *et al*., "Terahertz multi-spectral mueller matrix polarimetry on leaf using only orthogonal-polarization measurements," IEEE Trans. on Terahertz Sci. Technol. **11**(6), 609–619 (2021).
17. Z. B. Harris, K. Xu, and M. H. Arbab, "A handheld polarimetric imaging device and calibration technique for accurate mapping of terahertz stokes vectors," Sci. Reports **14**, 17714 (2024).
18. K. Xu, Z. B. Harris, and M. H. Arbab, "Polarimetric imaging of back-scattered terahertz speckle fields using a portable scanner," Opt. Express **31**(7), 11308–11319 (2023).
19. N. Gurjar, K. Bailey, and M. O. El-Shenawee, "Polarimetry terahertz imaging of human breast cancer surgical specimens," J. Med. Imaging **11**(6), 065503 (2024).
20. N. C. J. V. D. Valk, W. A. M. V. D. Marel, and P. C. M. Planken, "Terahertz polarization imaging," Opt. Lett. **30**(20), 2802–2804 (2005).
21. R. Zhang, Y. Cui, W. Sun, and Y. Zhang, "Polarization information for terahertz imaging," Appl. Opt. **47**(34), 6422–6427 (2008).
22. X. Wang, Y. Cui, W. Sun, *et al*., "Terahertz polarization real-time imaging based on balanced electro-optic detection," J. Opt. Soc. Am. A **27**(11), 2387–2393 (2010).
23. S. N. Lowry, L. E. Bergen, M. J. Herbert, *et al*., "Polarization-resolved terahertz time-domain imaging enabled by single pixel imaging," APL Photonics **10**, 076120 (2025).
24. K. Xu, Z. B. Harris, P. Vahey, and M. H. Arbab, "THz polarimetric imaging of carbon fiber-reinforced composites using the portable handheld spectral reflection (PHASR) scanner," Sensors **24**(23), 7467 (2024).
25. K. Xu and M. H. Arbab, "Terahertz polarimetric imaging of biological tissue: Monte carlo modeling of signal contrast mechanisms due to mie scattering," Biomed. Opt. Express **15**(4), 2328–2342 (2024).
26. E. Heller, K. Xu, Z. B. Harris, and M. H. Arbab, "Terahertz mie scattering in tissue: diffuse polarimetric imaging and monte carlo validation in highly attenuating media models," J. Biomed. Opt. **30**(6), 066001 (2025).

27. J. Baek, J. Kim, J. H. Seol, and M. Kim, "All-dielectric polarization-sensitive metasurface for terahertz polarimetric imaging," Sci. Reports **14**, 7544 (2024).
28. X. Li, J. Li, Y. Li, *et al*., "High-throughput terahertz imaging: progress and challenges," Light. Sci. & Appl. **12**, 233 (2023).
29. P. Carelli, F. Chiarello, S. Cibella, *et al*., "A fast terahertz spectrometer based on frequency selective surface filters," J. Infrared, Millimeter, Terahertz Waves **33**, 505–512 (2012).
30. R. Sebastian, R. Ahmad, J. Lafrenière-Greig, X. Ropagnol, and F. Blanchard, "Real-time material identification using a fast and simplified ai-assisted terahertz spectrometer," IEEE Trans. on Terahertz Sci. Technol. (2025).
31. R. Ahmad, J. Lafrenière-Greig, X. Ropagnol, and F. Blanchard, "Terahertz polarimetric system based on polarization sensitive frequency selective surfaces," Photonics North (PN), pp. 1–2 (2025).
32. R. Ahmad, X. Ropagnol, and F. Blanchard, "Polarization-sensitive terahertz frequency-selective surface based on a laser-cutting technique [Invited]," Journal of the Optical Society of America B **43**(3), A152-A163 (2026).
33. R. Ahmad, X. Ropagnol, N. D. Trinh, C. Bois, and F. Blanchard, "Reconfigurable screen-printed terahertz frequency selective surface based on metallic checkerboard pattern," Flex. Print. Electron. **9**(2), 025005 (2024).
34. E. Castro-Camus, J. Lloyd-Hughes, M. D. Fraser, H. H. Tan, C. Jagadish, and M. B. Johnston, "Detecting the full polarization state of terahertz transients," Terahertz and Gigahertz Electronics and Photonics V, 61200Q (2006).
35. L. Guiramand, J. Lafrenière-Greig, X. Ropagnol, and F. Blanchard, "Near-field terahertz electro-optical imaging based on a polarization image sensor," New J. Phys. **26**, 103007 (2024).
36. R. Ahmad, F. Blanchard and, R. Al Hadi, "Characterization and Polarization Sensitivity Analysis of CMOS Terahertz Detector," Photonics North 2025, pp. 1-2, (2025).
37. P. U. Jepsen, "Phase retrieval in terahertz time-domain measurements: A 'how to' tutorial," J. Infrared, Millimeter, Terahertz Waves **40**, 395–411 (2019).
38. R. Barakat, "Jones matrix equivalence theorems for polarization theory," Eur. J. Phys. **19**(3), 209–216 (1998).
39. M. Menzel, K. Michielsen, H. D. Raedt, *et al*., "A jones matrix formalism for simulating three-dimensional polarized light imaging of brain tissue," J. Royal Soc. Interface **12**(111), 20150734 (2015).
40. M. Neshat and N. P. Armitage, "Improved measurement of polarization state in terahertz polarization spectroscopy," Opt. Lett. **37**(11), 1811–1813 (2012).
41. S. Singh, M. Eisenmann, Y. Aso, and K. Somiya, "Complete birefringence and jones matrix characterization using arbitrary polarization," Opt. Express **33**(8), 17462–17476 (2025).
42. G. Ghosh, "Dispersion-equation coefficients for the refractive index and birefringence of calcite and quartz crystals" Optics Communications **163**(1-3), 95-102 (1999).